\documentclass{article}

\usepackage{arxiv}

\usepackage[utf8]{inputenc} 
\usepackage[T1]{fontenc}    
\usepackage{hyperref}       
\usepackage{url}            
\usepackage{booktabs}       
\usepackage{amsfonts}       
\usepackage{nicefrac}       
\usepackage{microtype}      
\usepackage{lipsum}
\usepackage{graphicx}
\usepackage{upgreek}
\usepackage{stfloats}
\usepackage{amsmath}

\title{Optical design of the laser launch telescope via physical optics theorem for Laser Guide Star Facility}

\author{Yan Mo, ZhengBo Zhu, Zichao Fan, Donglin Ma\thanks{Shenzhen Huazhong University of Science and Technology, Shenzhen 518057, China}   \\
  School of Optical and Electronic Information and Wuhan National Laboratory of Opto-electronics\\
  Huazhong University of Science and Technology\\
  Wuhan 430074, China \\
  \texttt{madonglin@hust.edu.cn} \\
}

\begin{document}
\maketitle

\begin{abstract}
The Laser Guide Star Facility (LGSF), as the most important part of the adaptive optics system of the large ground-based telescope, is aimed to generate multiple laser guide stars at the sodium layer. Laser Launch Telescope is employed to implement this requirement by projecting the Gauss beam to the sodium layer with a small beam size in LGSF system. As the diffraction and interference effects of laser’s long-distance transmission, the conventional optical design based on the geometrical optics mechanism cannot achieve the expected laser propagation. In this paper, we propose a method to design optical system for laser launch telescope based on the physical optics theorem to generate an acceptable light spot at the sodium layer in the atmosphere. Besides, a tolerance analysis method based on physical optics propagation is also demonstrated to be necessitated to optimize the system’s instrumentation performance. The numerical results show that the optical design considering physical optics propagation is highly rewarding and even necessitated in many occasions, especially for laser beam propagation systems. 
\end{abstract}

\keywords{The Laser Guide Star Facility \and Laser lunch telescope \and Optical design \and Physical optics propagation }

\section{Introduction}
The Laser Guide Star Facility (LGSF) is one of the most important parts of large ground-based telescopes to improve the capability in high-resolution imaging of faint stars. Specifically, it is used to generate artificial laser guide stars for adaptive optics (AO) systems to compensate for the perturbation caused by the atmosphere. The laser guide star technique was firstly put forward by Happer $et$ $al$ in 1982 \cite{happer1994atmospheric} and then the experiment was implemented by Primmerman $et$ $al$ \cite{primmerman1991compensation}. In 2001, the laser guide star AO system with laser guide star was firstly installed on Keck I \cite{chin2012keck} and later on Keck II \cite{chin2016keck}. With the continuous progress of astronomical optical technology, the laser guide star system projecting several asterisms was assembled on Gemini telescope in 2011 \cite{d2012gemini},  which demonstrated the well performance and high reliability of the LGSF. From then on, the LGSF has been widely used by many other famous observatories including VLT \cite{hackenberg2014assembly}, Subaru \cite{minowa2012subaru}, and Thirty Meter Telescope (TMT) \cite{li2018design}, etc.

In order to project and expend the laser beam into sodium layer, a crucial component called Laser Launch Telescope(LLT) is employed in LGSF system. As for the optical requirements, the most important goal of LLT is to eliminate aberration as much as possible to generate an acceptable light spot among the sodium layer at a predefined altitude (120km) and maintain a high ratio of encircled energy. Specifically, the RMS wavefront error (WFE) of the LLT design should be limited to a reasonable value. A high encircled energy ratio means a high energy utilization efficiency, which contributes to produce bright artificial laser guide stars.

As the long-distance of propagation of laser beam in free space, it is necessary to take the diffraction and interference effects into consideration in the optical design process of LLT. In other words, the optical design of LLT should be implemented in the physical optics theorem. However, it is almost impossible to conduct the optical design for LLT directly based on the physical optics theory simply relying on commercial optical design software such as ZEMAX. As a rule of thumb, an optimal design process is to design the initial optical structure with the assumption of geometrical optics approximation , and then optimize it based on the physical optics theorem.

In this paper, we propose a method to design optical system for LLT  based on the physical optics theorem. In this method, we firstly analyze the physical propagation model of laser beam, then design the initial optical system of LLT based on the geometrical optics assumption. After that, the optical performance of the initial LLT is evaluated based on physical optics. The RMS wavefront error (WFE) and the encircled energy ratio are selected as the criteria for evaluation of optical performance. Next, the optical system is updated with the physical optics theorem to achieve the predefined optical requirements. Finally, we provide a tolerance analysis method based on Gauss beam propagation to predict the expected optical performance of LLT with instrumentation errors. For the finally obtained optical system, the encircled energy ratio within a diameter of 233mm at 120km exceeds $92.5\%$ considering the tolerance allocation. And the largest RMS WFE is less than 0.016$\lambda$ with the working temperature ranging from -5$^{\circ}$C to 20$^{\circ}$C.

\section{Design Considerations}
\label{sec:examples}
\begin{figure}
\centering\includegraphics[width=8cm]{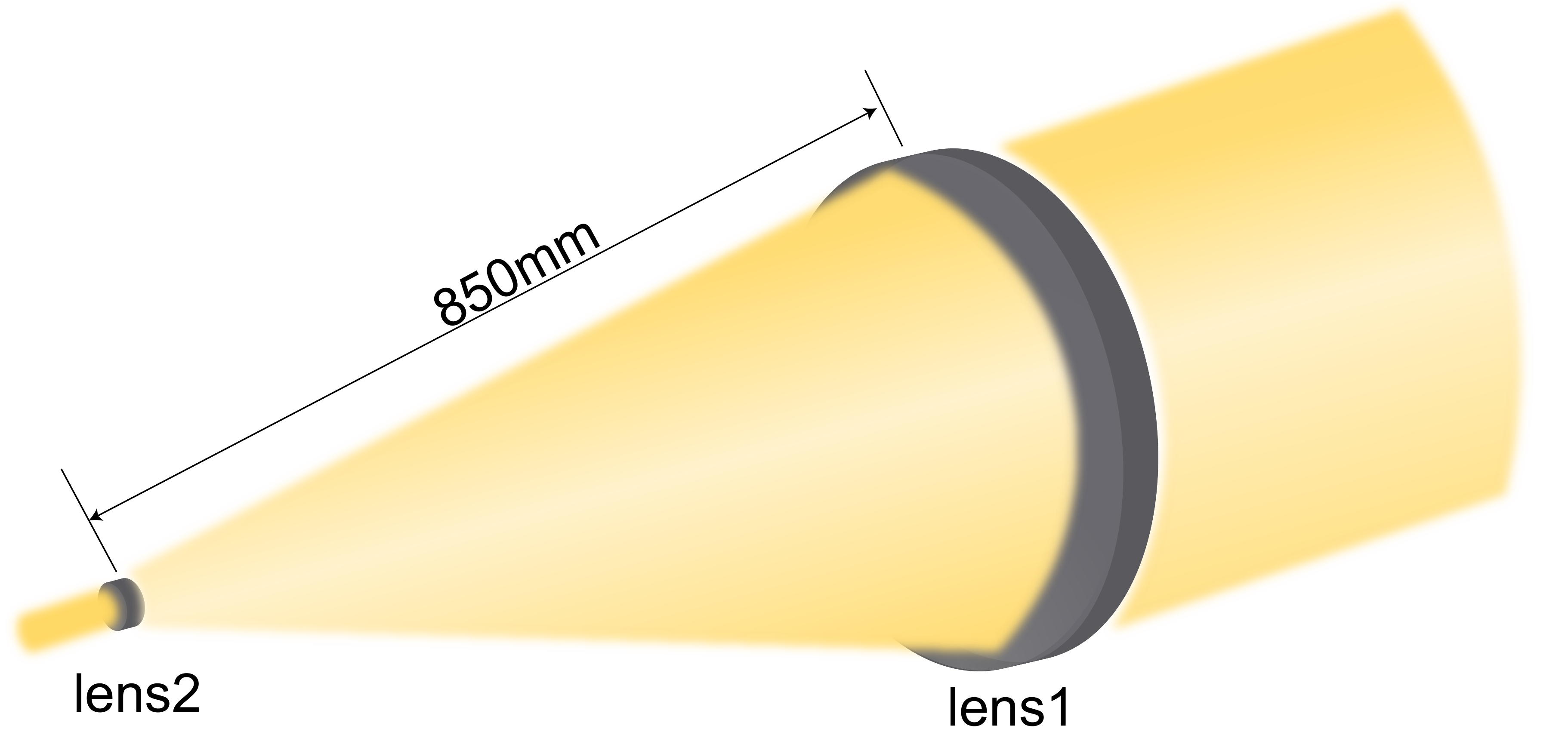}
\caption{Optical layout of LLT}
\label{fig:1}
\end{figure}

The LLT is a laser beam expander essentially. To  avoid  the  extremely  tight  optical  and  mechanical  tolerances, we choose the Galilei telescope as the initial structure of LLT. Additional advantages of this choice are the compact configuration and the avoidance of the internal focal point, which is beneficial to the mechanical fabrication. Only a single working wavelength (589nm) is considered, and a Galilei telescope configuration with two singlets is employed.  The optical layout of LLT is shown in Fig. \ref{fig:1}. 

For the general applications, the working temperature ranges from -5$^{\circ}$C to 20$^{\circ}$C is considered. To reduce the difficulty of mechanical alignment, the distance between Lens 1 and Lens 2 is set as 850mm in consideration of the general demands. Besides, the distance between Lens 1 and Lens 2 is expected to be adjustable within a range of ±0.25 mm to compensate the performance degradation due to the manufacturing and assembly errors as well as the environmental disturbance. We choose the field of view (FOV) as 0.06$^{\circ}$ to match the general AO systems. As mentioned above, the main goal of LLT is to produce a small light spot at the sodium layer while maintaining high energy efficiency as much as possible, we select the RMS WFE and the enclosed energy ratio as the criteria for the evaluation of optical performance. Based on the science requirement of general AO systems, the radius of the light spot is usually limited to 233mm at 120km in altitude. The design specifications are expressly presented in Table \ref{tab:T1}. 

\begin{table}[htbp]
\centering
\caption{\bf Design Requirements for LLT}
\begin{tabular}{cc}
\hline
Parameters & Value \\
\hline
Wavelength & 589nm \\
Pupil\ position & Lens\ 1\\
RMS\ WFE & 0.037$\lambda$\\
FOV & 0.06$^{\circ}$\\
Working\ temperature & -5$^{\circ}$C\ $\sim$\ 20$^{\circ}$C\\
Distance\ between\ Lens\ 1\ and\ Lens\ 2 & 850 $\pm$ 0.25mm\\
Radius\ of\ light\ spot & <233mm\\
Encircled\ energy\ ratio & >$90\%$\\
\hline
\end{tabular}
  \label{tab:T1}
\end{table}

\section{Optical Design}
\subsection{The Geometrical Optical Design}

The spherical aberration is the main error source that contributes the  enclosed energy loss in the sodium layer. Based on Seidel sums \cite{gross29handbook}, all of the primary aberrations coefficients of the given optical system can be numerically calculated. For the spherical aberration:

\begin{equation}
{{S}_{1}}=\frac{1}{4}{{\phi }^{3}}{{y}^{3}}(A{{X}^{2}}-BXM+C{{M}^{2}}+D)
\label{eq:1}
\end{equation}

\begin{equation}
{{W}_{040}}\text{=}\frac{1}{8}{{S}_{1}}
\label{eq:2}
\end{equation}
where $\phi$ is the  refractive power, $M$ represents the position or conjugate parameter, $X$ denotes the bending parameter and $A$, $B$, $C$, $D$ are constants related to the refractive index. The position or conjugate parameter $M$ is given by:
\begin{equation}
M=\frac{{u}'+u}{{u}'-u}=\frac{1+m}{1-m}
\label{eq:3}
\end{equation}
where $u$ and $u'$ are the paraxial marginal ray angle before and after the lens respectively, $m$ stands for the magnification. The bending parameter $X$ is determined by:
\begin{equation}
{X}=\frac{{{c}_{1}}+{{c}_{2}}}{{{c}_{1}}-{{c}_{2}}}
\label{eq:4}
\end{equation}
where $c_1$ and $c_2$ are curvatures of a lens. From Eq. (\ref{eq:1}), it is obvious that the spherical aberration $S_1$ depends on the square of the bending parameter $X$. Therefore a suitable choice of bending allocation for two lenses is necessitated to minimize spherical aberration. During the optimization progress, the location of the beam waist is constrained. The optimized optical parameters are listed in Table \ref{tab:T2}. 

\begin{figure}
\centering\includegraphics[width=8cm]{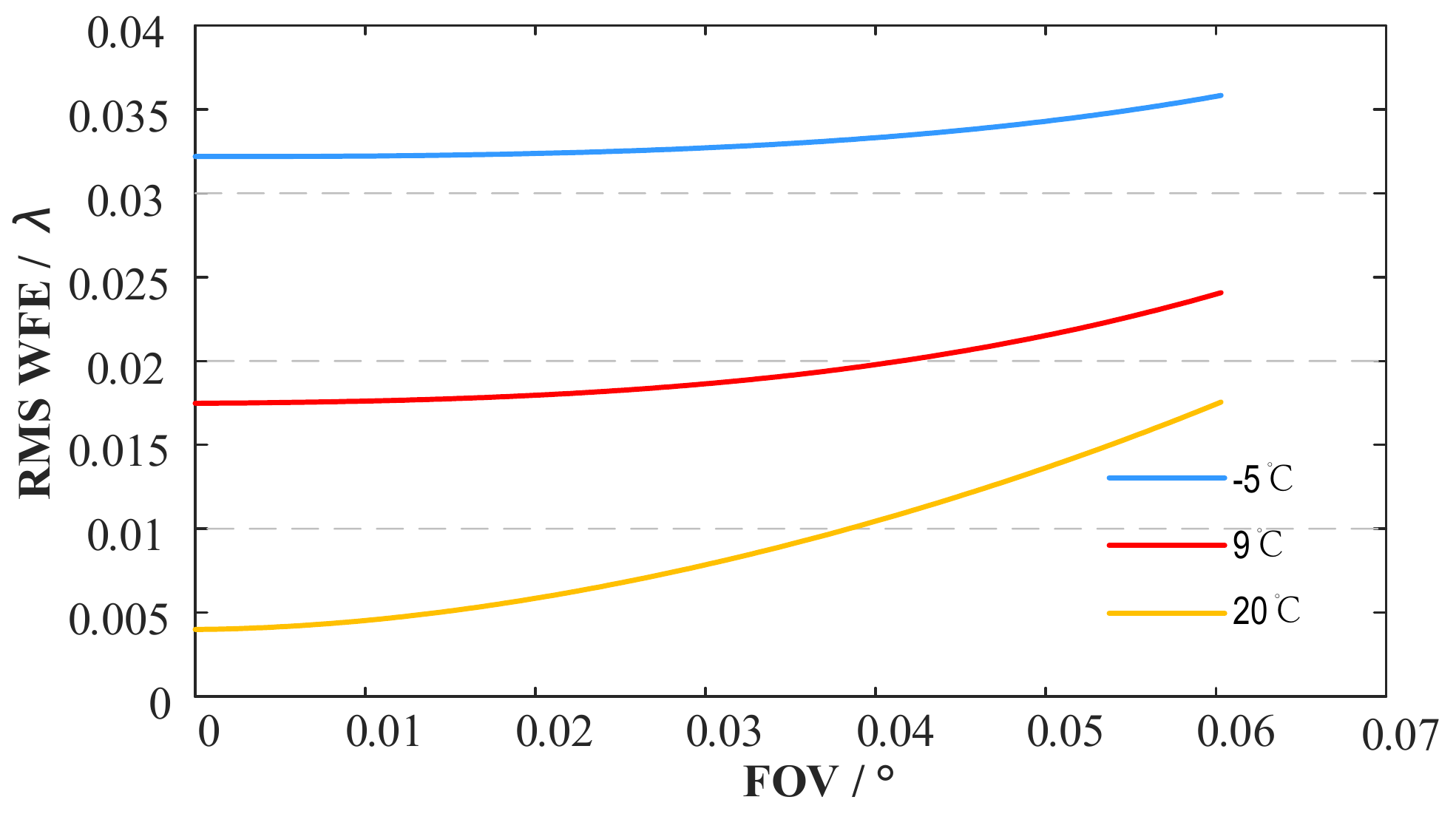}
\caption{RMS WFE Vs FOV for different temperature}
\label{fig:2}
\end{figure}

To evaluate the optical performance of the obtained optical system at different temperatures, the thermal analysis is implemented. Three different working temperatures including -5$^{\circ}$C, 9$^{\circ}$C and 20$^{\circ}$C are considered. The distance between Lens 1 and Lens 2 is selected as the compensator.  As  illustrated  in Fig. \ref{fig:2}, the initial design shows a good optical performance under different working conditions and meets the requirement of RMS WFE over full FOV. However, the compensation distance reaches 0.5134mm, which may not satisfy the mechanical constraint.

\begin{table}[htbp]
\centering
\caption{\bf Optical parameter for LLT initial design}
\begin{tabular}{ccccc}
\hline
Element & Material & Curvature & Thickness & Conic\\
 & &  Radius(mm) & (mm) & \\
\hline
Lens1-S1 & SILICA & 397.239 & 70.022 & -0.411 \\
Lens1-S2 &  & 1641.119 & 920.000 &  \\
Lens2-S1 & SILICA & -68.700 & 15.236 & -0.972 \\
Lens2-S2 &  & 1120.145 &  &  \\
\hline
\end{tabular}
  \label{tab:T2}
\end{table}

\subsection{Physical Optics Performance}
\begin{figure}
\centering\includegraphics[width=8cm]{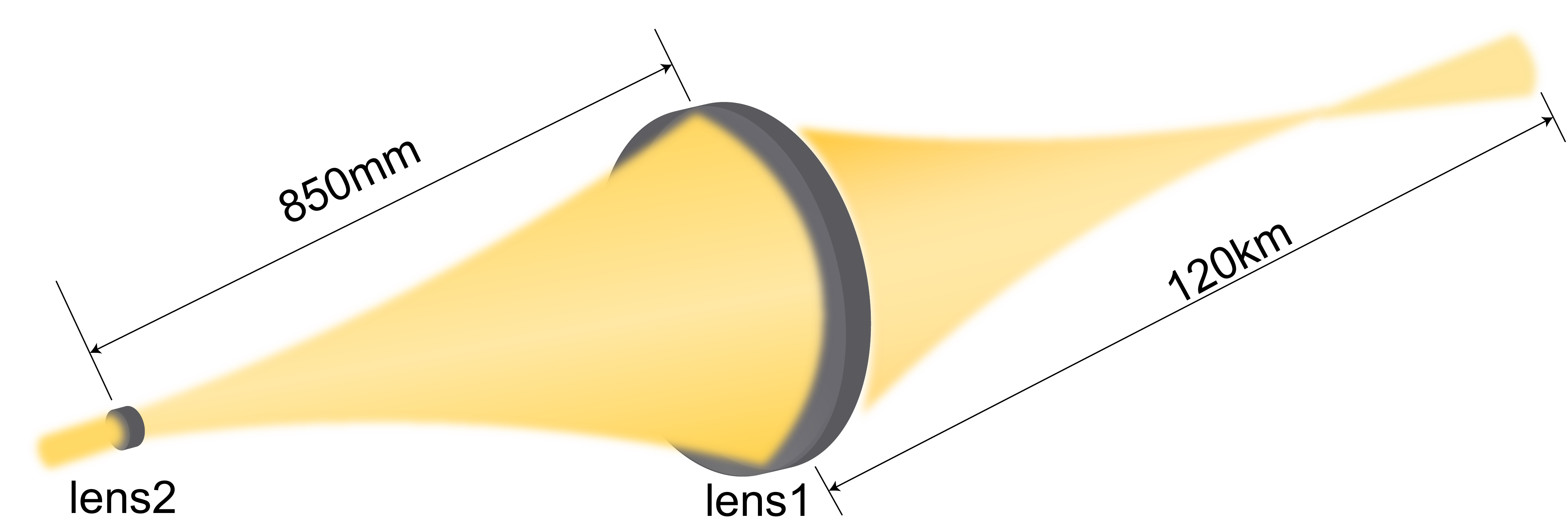}
\caption{Gauss Beam propagation of LLT}
\label{fig:3}
\end{figure}

Note that, it is impossible to directly project the beam waist to the sodium layer by the above obtained optical system according to the relationship between beam waist location and waist size, which is given by
\begin{equation}
w={{w}_{0}}^{2}\sqrt{(1+\frac{{{z}^{2}}}{{{f}^{2}}})}
\label{eq:5}
\end{equation}
where $w_0$ is the semi-diameter of the waist, $z$ represents the propagation distance, and $f$ demotes the rayleigh range. The output beam $1/e^2$ diameter is set as 240mm to avoid beam clipping \cite{holzlohner2008physical} referring to the existing designs \cite{bonaccini2003vlt, C2002Gemini, boyer2010tmt}. Based on the simulation result, the optimum beam waist locates at 34km in altitude with a waist radius of 100mm. As the long-distance propagation of laser beam in free space, the diffraction and interference effects can not be ignored. To precisely evaluate the optical performance of this specific optical system, the physical propagation model of laser should be analyzed. The Gauss beam propagation model based on the physical simulation of the above obtained LLT system is shown in Fig. \ref{fig:3}.

As the brightness and the beam quality is highly required by the AO System, the encircled energy ratio inside a specific radius is chosen as one of the most important assessment criteria of optical performance. The amplitude of an ideal collimated Gauss beam can be represented by:
\begin{equation}
A(r)={{a}_{0}}\exp (\frac{-{{r}^{2}}}{w}).
\label{eq:6}
\end{equation}

And the corresponding irradiance is calculated as:
\begin{equation}
I(r)={{I}_{0}}\exp (\frac{-2{{r}^{2}}}{w})
\label{eq:7}
\end{equation}
where $r$ denotes the light spot radius and $w$ represents the specific value of $r$ when the irradiance equals $I_0/e^2$.
As expressed in Eq. (\ref{eq:7}), the beam brightness and quality is relevant to $w$, which is given by Eq. (\ref{eq:5}). Therefore, the ratio of the encircled energy of an ideal Gauss beam can be calculated as:
\begin{equation}
E(r=a)=\frac{\int_{0}^{a}{\int_{0}^{2\pi }{I(r)2\pi rdrd\theta }}}{\int_{0}^{\infty }{\int_{0}^{2\pi }{I(r)2\pi rdrd\theta }}}
\label{eq:8}
\end{equation}

\begin{figure}
\centering\includegraphics[width=8cm]{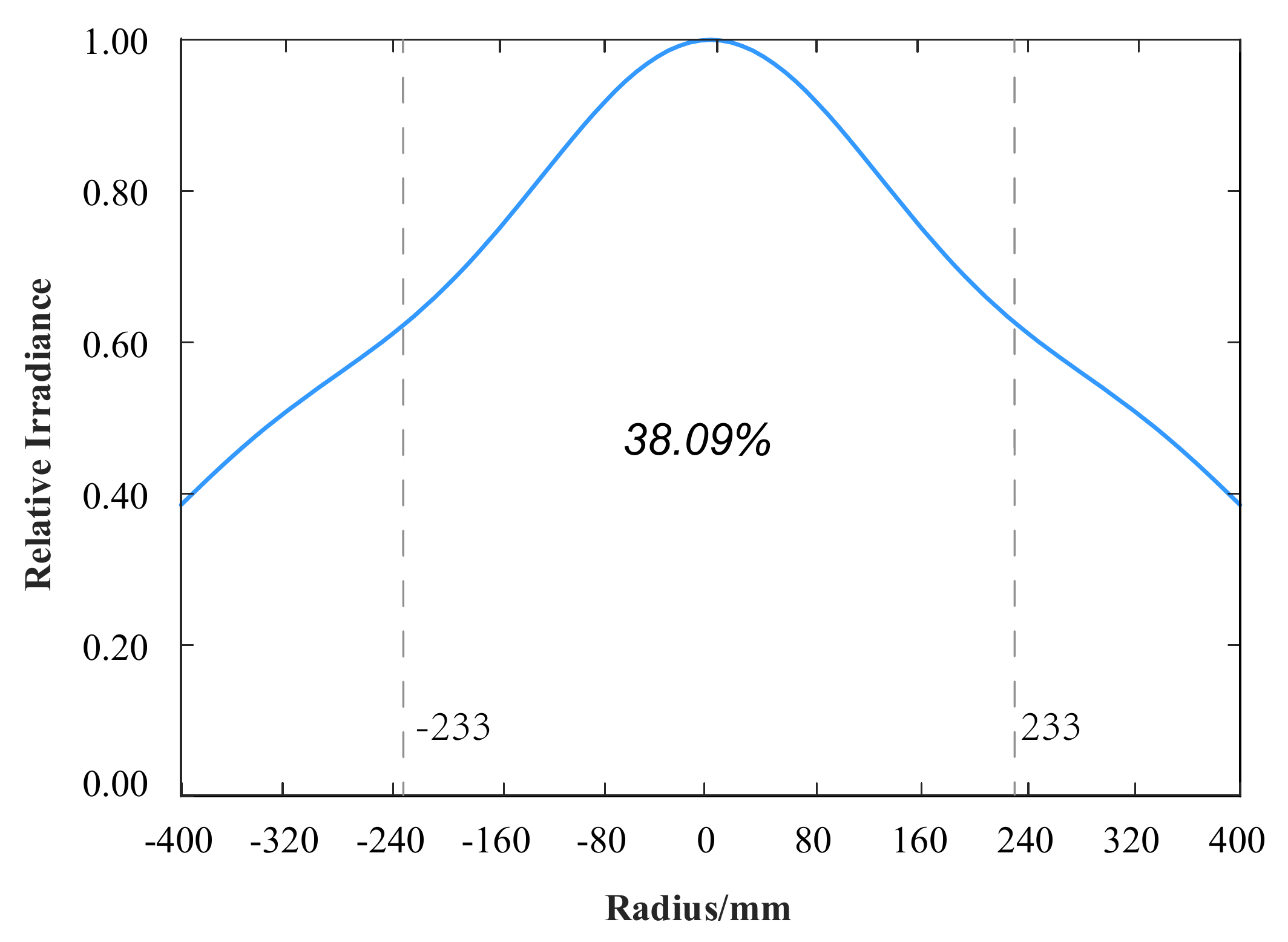}
\caption{The encircled energy ratio of the initial design}
\label{fig:4}
\end{figure}

The normalized irradiance distribution at 120km is shown in Fig. \ref{fig:4}. Only $38.09\%$ of energy is encircled inside the circular domain with a radius of 233mm, which cannot satisfy the science requirement as listed in Table \ref{tab:T1}.  This result indicates that the optical design optimization procedure under geometrical optics evaluation criterion is ineffective for the long-distance propagation of laser beam. A precise optical design is necessary, for instance, the optical design based on the physical optics theorem is needed.

\subsection{Update The Optical Design Using Physical Optics Theorem}
In Section 3.A, the location of the beam waist is constrained in the optimization process while minimizing the RMS WFE, however, the optical performance can not meet the requirements as analyzed in Section 3.B. In this Section, the merit function is replaced by the physical optics evaluation criterion. Meanwhile, the geometrical ray tracing performance should be satisfied.

\begin{figure}
\centering\includegraphics[width=8cm]{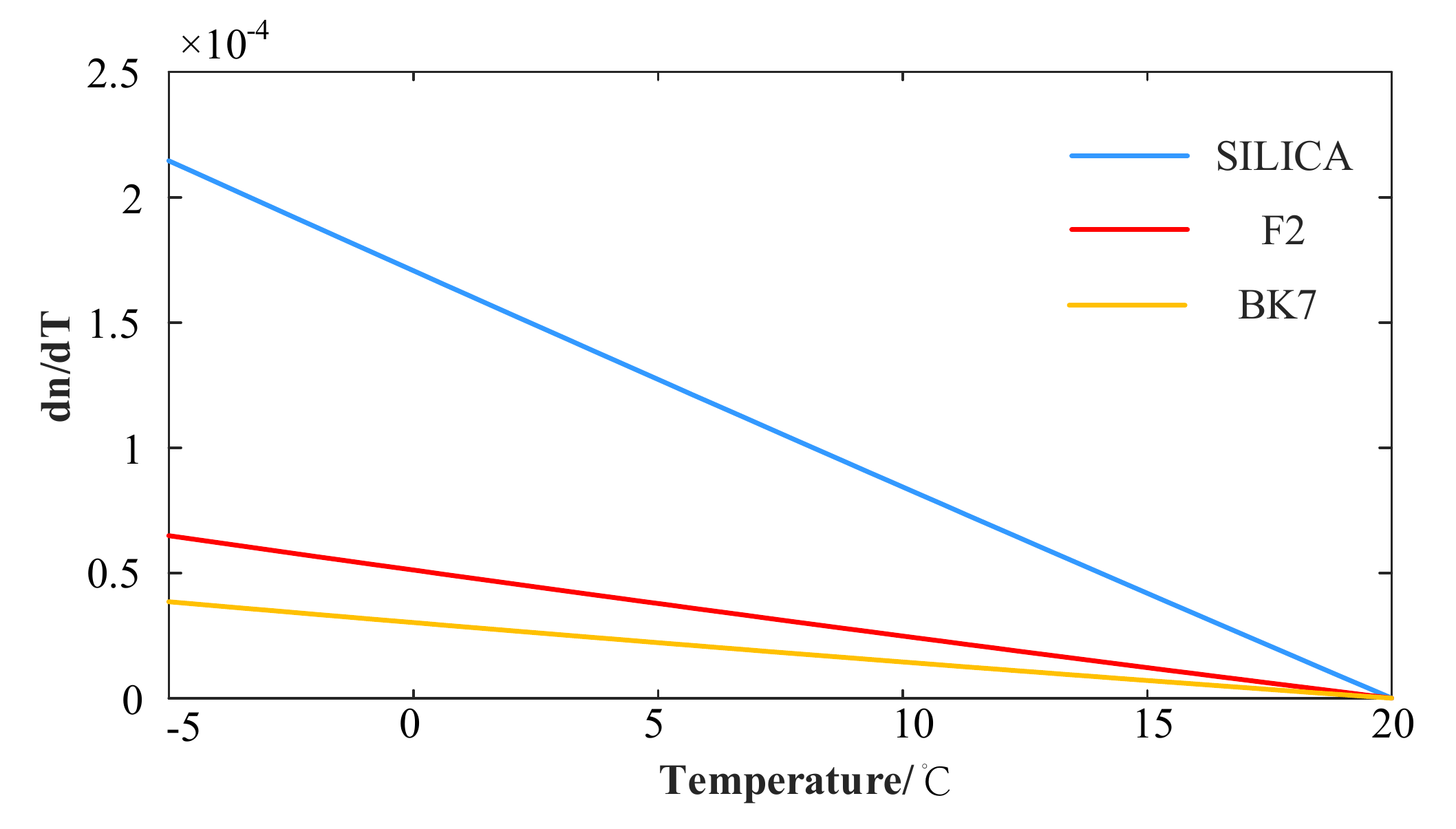}
\caption{Temperature coefficient of the absolute refractive index of three different optical glasses}
\label{fig:5}
\end{figure}

\begin{figure}
\centering\includegraphics[width=8cm]{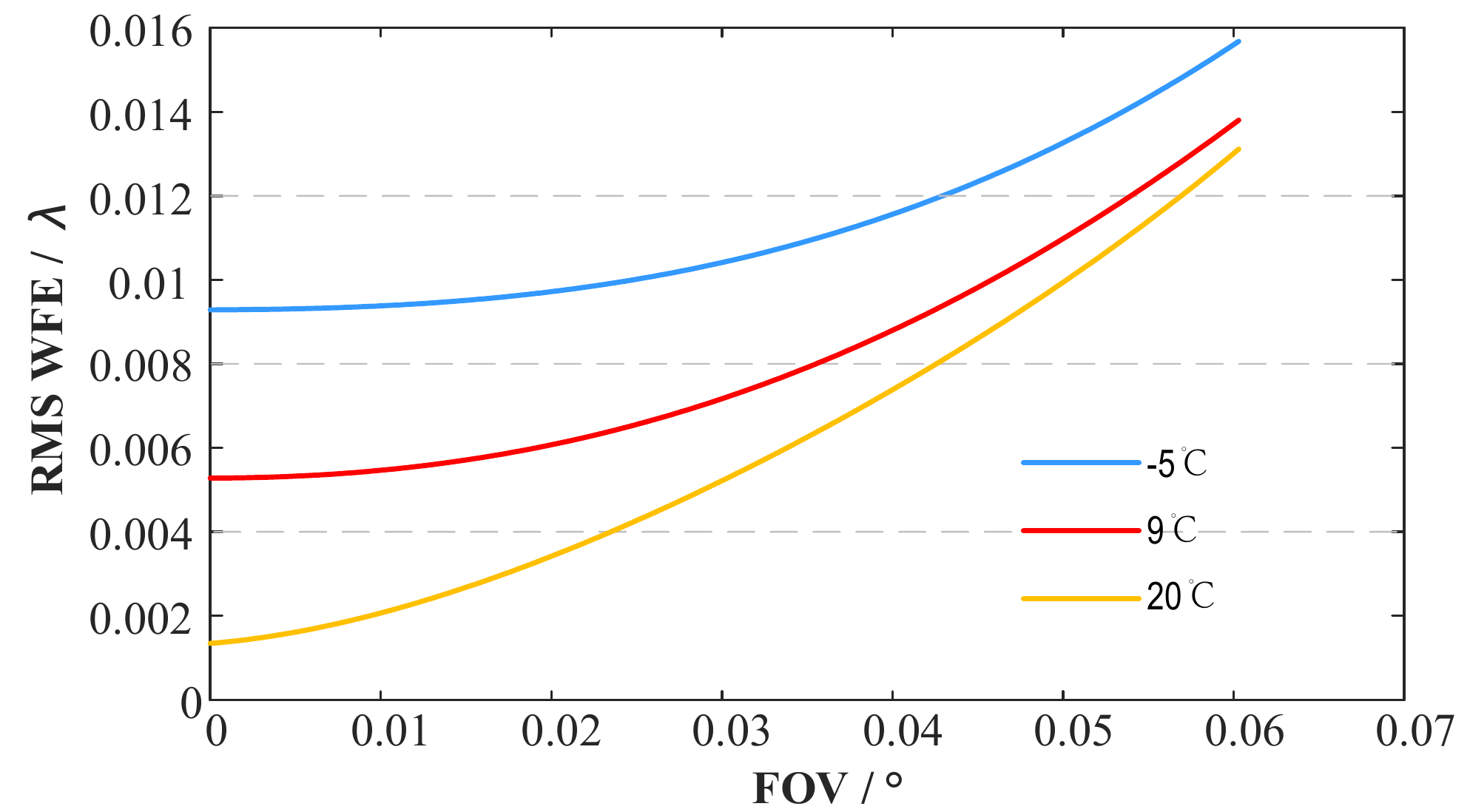}
\caption{RMS WFE vs Field of View under different temperatures}
\label{fig:6}
\end{figure}

As mentioned above, a shorter compensation distance is preferred considering the instrumentation, and we make a trade-off discussion of material choice to minimize the compensation distance. Three different kinds of typical optical glass including SILICA, F2, and BK7 are taken into consideration. The temperature coefficient of the absolute refractive index for a specific material can be expressed:

\begin{equation}
\frac{dn}{dT}=\frac{{n}^{2}-{1}}{2n}({D}_{0}+{2}{D}_{1}{\Delta}{T}+{3}{D}_{2}{\Delta}{T}^{2}+\frac{{E}_{0}+{2}{E}_{1}{\Delta}{T}}{{\lambda}^{2}-{\lambda}_{TK}^{2}}
\label{eq:9}
\end{equation}
where $n$ represents the refractive index relative to vacuum; $\Delta T$ is the temperature difference; $\lambda$ stands for the wavelength; $D_0$, $D_1$, $D_2$, $E_0$, $E_1$, and $\lambda_{TK}$ are constants depending on the glass type. The temperature coefficient of the absolute refractive index of three different types of optical glass is shown in Fig. \ref{fig:5}. It is obvious that F2 and BK7 has a relatively lower temperature coefficient of refractive index compared with SILICA, and we choose F2 and BK7 as the new materials for LLT and conduct the optimization with the physical optics evaluation criterion on the initial design. The distance between Lens 1 and Lens 2 is selected as a compensator. The final positions of the optical components after the optimization process are presented in Table \ref{tab:T3}. The optical performance is also evaluated by RMS WFE and the encircled energy ratio with different working temperatures. The relationship between RMS WFE and FOV is shown in Fig. \ref{fig:6}. It shows that the optimized design shows good optical performance in terms of RMS WFE and satisfies the predefined specifications well. The largest RMS WFE is less than 0.016$\lambda$ in the temperature range of -5$^{\circ}$C to 20$^{\circ}$C.  The compensation distance between Lens 1 and Lens 2 is reduced to 0.033mm, which is far less than that of the initial design.

\begin{table}[htbp]
\centering
\caption{\bf Optical parameter for LLT initial design}
\begin{tabular}{ccccc}
\hline
Element & Material & Curvature & Thickness & Conic\\
 & &  Radius(mm) & (mm) & \\
\hline
Lens1-S1 & BK7 & 392.831 & 70.000 & -0.378 \\
Lens1-S2 &  & 1356.645 & 850.000 &  \\
Lens2-S1 & F2 & -82.124 & 13.000 & -0.824 \\
Lens2-S2 &  & 6546.516 &  &  \\
\hline
\end{tabular}
  \label{tab:T3}
\end{table}

\begin{figure}
\centering\includegraphics[width=8cm]{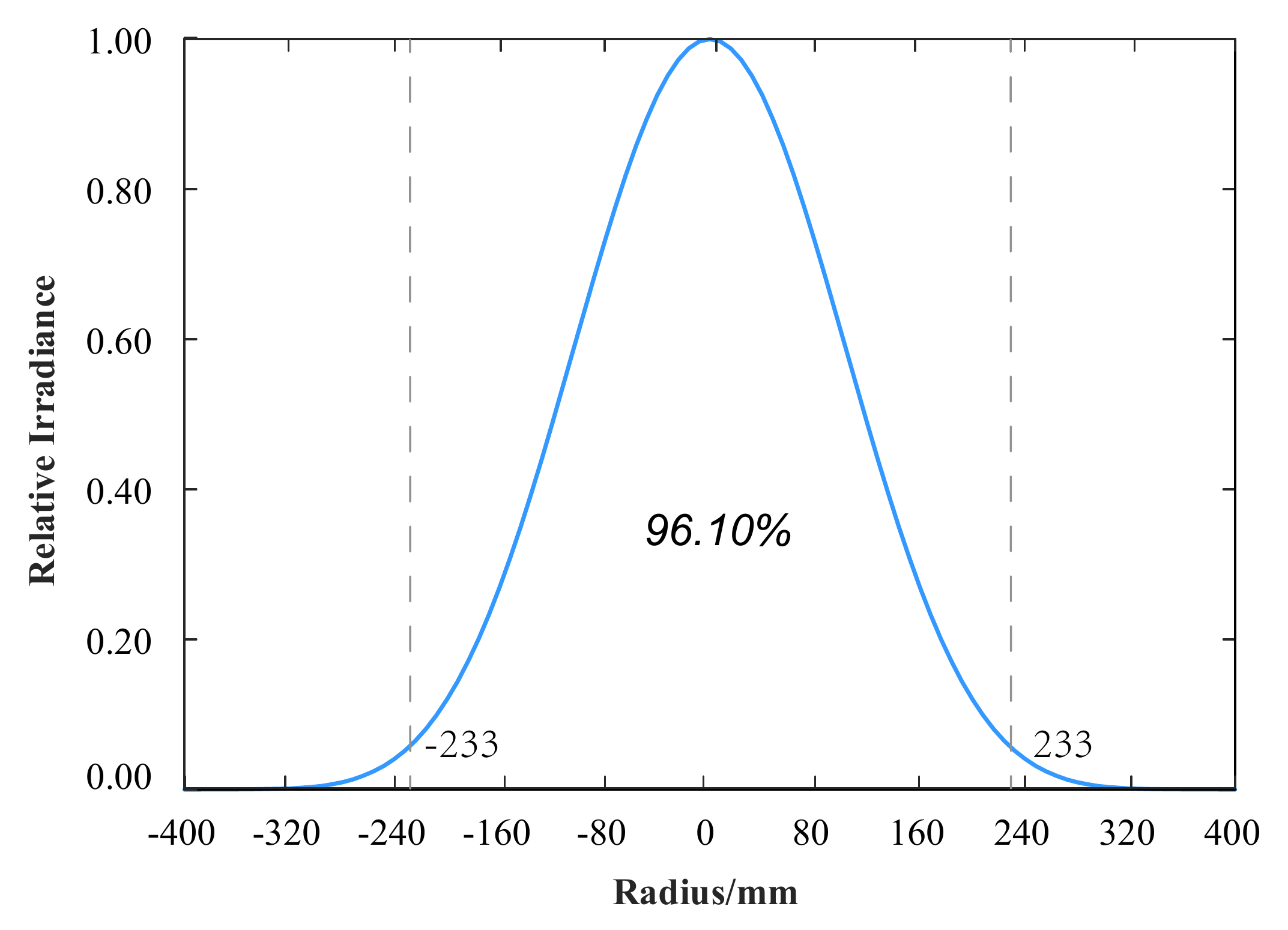}
\caption{The encircled energy ratio of the optimized design.}
\label{fig:7}
\end{figure}

 Similarly, we evaluate the encircled energy ratio within a diameter of 233mm at 120km in altitude, as shown in Fig. \ref{fig:7},  wherein 96.10$\%$ energy efficiency is achieved. All of these results demonstrate the optical design based on the physical optics theorem is necessitated for the laser's long-distance propagation after the initial design is constructed based on geometrical ray tracing.

\section{Tolerance Analysis}
Due to the special working environment of LLT, it is necessary to analyze its optical performance in different situations. In this research, we choose RMS WFE as the performance metric to allocate the tolerance. Assigning wavefront errors to each optical element of the LGSF system will be used to guide the fabrication of each optical component and the design of mechanical devices. The final RMS WFE budget for LLT should be limited to 0.037$\lambda$. A reasonable tolerance allocation of optical parameters is given after the sensitivity being analyzed. The estimated RMS WFE shows that the allocation can confirm to the performance deviation requirement well. As we discussed in Section 3, optical simulation based on the physical optics model is much more convinced than geometrical ray tracing. Hence, during the tolerancing process, we take the encircled energy ratio within a diameter of 233mm at 120km as the merit function in Monte Carlo tolerance analysis to estimate the expected physical optics propagation performance of LLT based on the allocation that we made above.

Firstly, the tolerance parameters need to be assigned. Each of the design parameters is perturbed within the range of tolerance allocation following a modified Gaussian normal distribution, which is given by: 

\begin{equation}
p(t)=\frac{1}{{\sqrt{{2}{\pi}{\sigma}}}}{\exp}({-\frac{{t}^{2}}{{2}{\sigma}^{2}}}).
\label{eq:10}
\end{equation}

\begin{figure}
\centering\includegraphics[width=8cm]{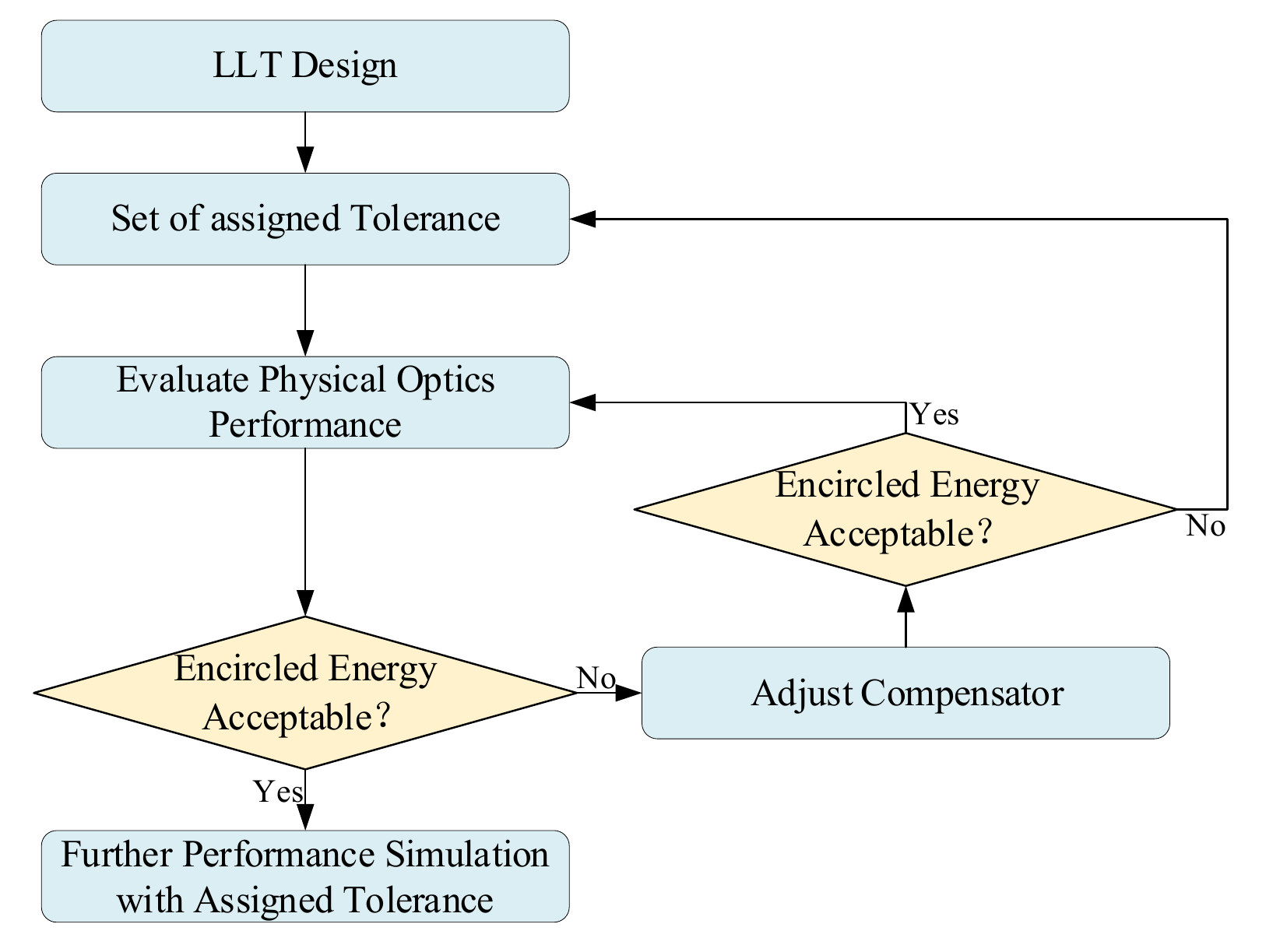}
\caption{Tolerance analysis flow chart.}
\label{fig:8}
\end{figure}

After the tolerance being assigned, a perturbed LLT module is created to evaluate the physical optics performance. If the enclosed energy of the perturbed module is unacceptable, a changeable parameter will be selected as a compensator. Hence, we perform the optimization of the distance between Lens 1 and Lens 2 of the perturbed module to compensate for the energy loss caused by the allocated tolerance parameter via the physical optics propagation simulation. During the optimization process, the compensation distance is restricted to 0.5mm. If the physical optics performance of the compensated module is acceptable, the enclosed energy ratio will be output as the result of each tolerance iteration. If not, the tolerance parameters are considered tight, and the tolerance iteration will be performed again until the enclosed energy ratio is acceptable. The tolerance process is shown in Fig. \ref{fig:8}.

\begin{figure}
\centering\includegraphics[width=8cm]{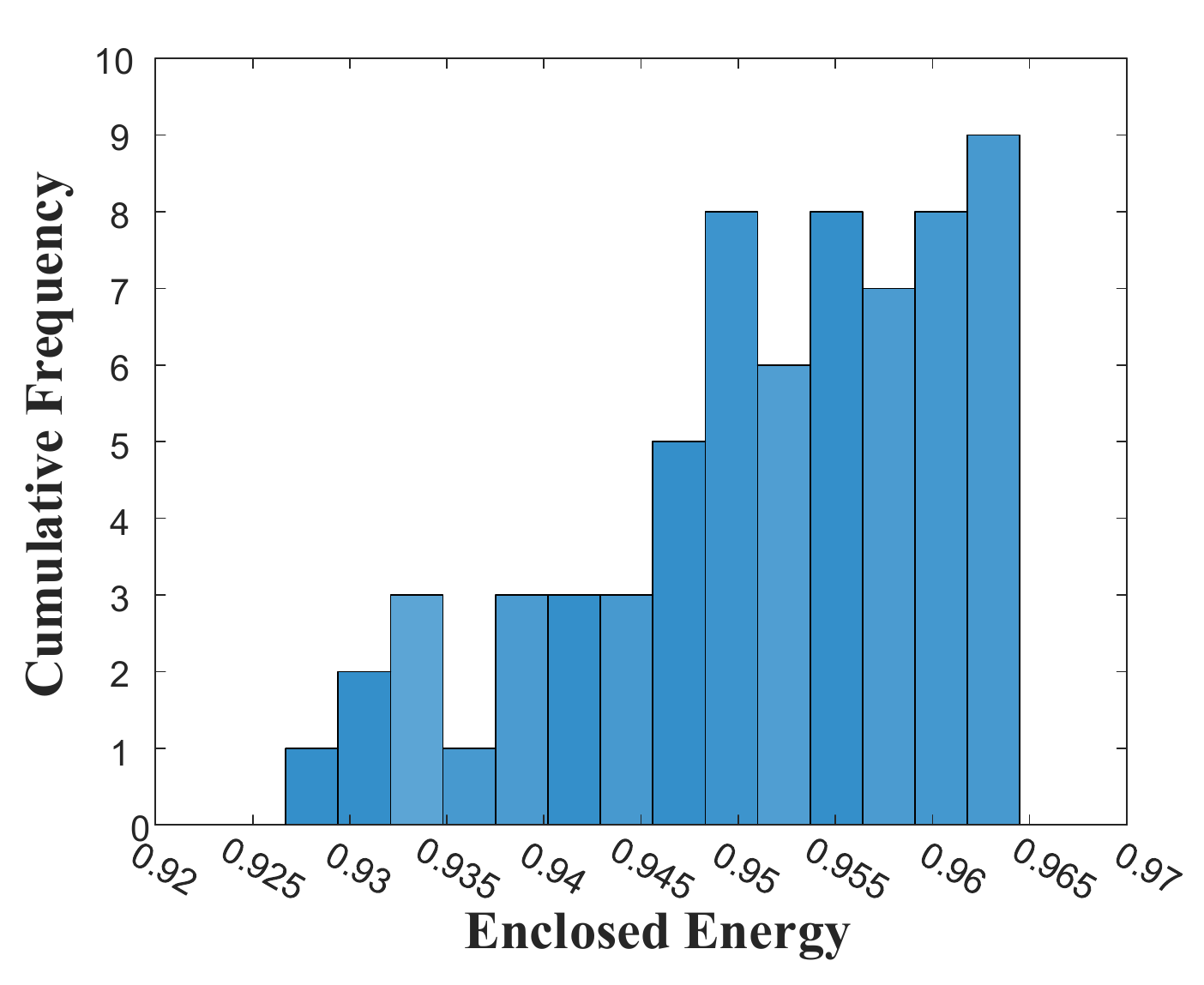}
\caption{Monte Carlo tolerance result.}
\label{fig:9}
\end{figure}

This process has been implemented in each iteration to guarantee a proper compensation distance and tolerance allocation. After 400 times of Monte Carlo analysis, a resealable tolerance allocation has been obtained as listed in Table \ref{tab:T4} considering the achievable fabrication ability, and parts of the Monte Carlo results is shown in Fig. \ref{fig:9}. It is obvious that the optimized system is not sensitive to fabrication when tolerance allocation is being considered. The enclosed energy ratio is larger than $92.5\%$ for all working conditions, which demonstrates that the provided tolerance method and allocation are convinced.

\begin{table}[htbp]
\centering
\caption{\bf Design Requirements for LLT}
\begin{tabular}{cc}
\hline
Parameters (mm) & Tolerance (mm)\\
\hline
Lens1\ surface1\ Radius & 0.1 \\
Lens1\ surface2\ Radius & 0.4 \\
Lens1\ thickness & 0.2\\
Lens2\ surface1\ Radius & 0.1 \\
Lens2\ surface2\ Radius & 0.4 \\
Lens2\ thickness & 0.1\\
\hline
\end{tabular}
  \label{tab:T4}
\end{table}

\section{Conclusion}
To summarize, we have provided a recommended optical design method for laser launch telescope  based on the physical optics theorem which is aimed to generate multiple laser guide stars at sodium layer. The optical design starts with the initial optical system design based on the geometrical optics assumption, and then we optimize the optical system via the physical optics theorem. Besides, the tolerance analysis is also provided to evaluate the feasibility of instrumentation based on the physics optics propagation.
The simulation results show that the proposed optical design method based on precise physical optics propagation is highly rewarding and even necessitated for the laser beam propagation systems. We believe that our work might provide a good guidance for researchers to design similar laser propagation systems in the future.

\bibliography{sample}



\bibliographystyle{unsrt}  
\end{document}